\documentstyle[preprint,aps,eqsecnum,epsf,floats]{revtex}

\newcommand{\PSbox}[3]{\mbox{\rule{0in}{#3}\includegraphics{#1}\hspace{#2}}}
\def\MgM{M\~gM}

\def\qslash{\not{\hbox{\kern-2pt $q$}}}
\def\delslash{\not{\hbox{\kern-2pt $\partial$}}}

\def\beq{\begin{equation}}
\def\eeq{\end{equation}}
\def\eeq{\end{equation}}
\def\bea{\begin{eqnarray}}
\def\eea{\end{eqnarray}}
\def\bq{\begin{quote}}
\def\eq{\end{quote}}
\def\lessim{\mathrel{\mathpalette\vereq<}}
\def\gtrsim{\mathrel{\mathpalette\vereq>}}
\def\lsim{\mathrel{\lesssim}}
\def\gsim{\mathrel{\gtrsim}}

\makeatletter
\def\vereq#1#2{\lower3pt\vbox{\baselineskip1.5pt \lineskip1.5pt
\ialign{$\m@th#1\hfill##\hfil$\crcr#2\crcr\sim\crcr}}}
\makeatother
\begin{document}
\preprint{\vbox{\hbox{SLAC-PUB-8325}
		\hbox{UCSD/PTH-00-01}
                \hbox{hep-ph/0001172}}}

\title{\vbox{\vskip 0.truecm} Minimal Gaugino Mediation}

\author{\vbox{\vskip 0.truecm} Martin Schmaltz $^x$ and Witold Skiba $^y$}
\address{
          \vbox{\vskip 0.truecm}
          $^x$SLAC, Stanford University, Stanford, CA 94309 \\
          {\tt schmaltz@slac.stanford.edu} \\
          \vbox{\vskip 0.truecm}
         $^y$Department of Physics, University of California at San Diego, 
          La Jolla,  CA 92093 \\
          {\tt skiba@einstein.ucsd.edu} }
\maketitle

\begin{abstract}%

We propose Minimal Gaugino Mediation as the simplest known solution to the supersymmetric
flavor and CP problems.
The framework predicts a very minimal structure for the soft parameters at
ultra-high energies: gaugino masses are unified and non-vanishing whereas
all other soft supersymmetry breaking parameters vanish. We show that this
boundary condition naturally arises from a small extra dimension and present
a complete model which includes a new extra-dimensional solution to the $\mu$ problem.
We briefly discuss the predicted superpartner spectrum as a function of the
two parameters of the model. The commonly ignored renormalization group evolution
above the GUT scale is crucial to the viability of Minimal Gaugino Mediation
but does not introduce new model dependence.

\end{abstract}

\newpage

%%%%%%%%%%%%%%%%%%%%%%
\section{Introduction}
%%%%%%%%%%%%%%%%%%%%%%

Hidden sectors are an essential ingredient of simple and natural models
of supersymmetry (SUSY) breaking. The unifying idea is that SUSY is
assumed to be broken in the hidden sector and then communicated to the
minimal supersymmetric standard model (MSSM) by messenger interactions
which are flavor blind. This structure results in flavor-universal
scalar masses and solves the SUSY flavor problem.

In the context of extra dimensions with branes such
hidden sectors are very natural.
If -- for example -- the MSSM  and the SUSY breaking
sector are confined to two different
parallel 3-branes embedded in extra dimensions, then the separation in
the extra dimensions forbids direct local couplings 
between the ``visible'' MSSM fields and the hidden sector.
However, fields on separated branes can still communicate
by exchanging bulk messenger fields. Couplings which arise from
such non-local bulk mode exchange are suppressed.
For a messenger of mass $M$ and brane separation $L$ the suppression
factor is $e^{-ML}$, which is the Yukawa propagator of the messenger
field exchanged between the two branes.

This suggest a very simple scenario for communicating SUSY breaking
to the MSSM which guarantees flavor-universal scalar masses.
If all light bulk fields have flavor blind couplings then the
soft SUSY breaking parameters generated by exchange of these messengers
preserve flavor. Heavy bulk modes may violate flavor maximally
but the resulting non-universal contributions to the scalar masses are
exponentially suppressed \cite{RS0}.

The two obvious candidates for bulk fields which can communicate
SUSY breaking to the Standard Model fields in a flavor-blind way
are gravity and the Standard Model gauge fields. Gravity as a bulk messenger
(``Anomaly Mediation'' \cite{RS0,GLMR}) leads to a very simple and
predictive model which unfortunately predicts negative slepton masses
and is therefore ruled out in its simplest and most elegant form.%
\footnote{For models which cure Anomaly Mediation
by introducing new fields and interactions see \cite{fixanomaly}.}
The alternative, Standard Model gauge fields as messengers (``Gaugino
Mediation''), has been proposed recently by D.E. Kaplan, Kribs,
and Schmaltz \cite{KKS}
as well as by Chacko, Luty, Nelson, and Ponton \cite{CLNP} and was found to
work perfectly. In Gaugino Mediation the MSSM matter fields
(quarks, leptons and superpartners) live on a ``matter brane'', 
while SUSY breaks on a parallel ``SUSY breaking brane'', and the MSSM gauge
superfields live in the bulk. Because the gaugino fields are bulk fields they
couple directly to the SUSY breaking and obtain soft masses.
The MSSM scalars are separated from SUSY breaking by the distance
$L$ and therefore obtain much smaller masses from non-local loops with
high momentum modes of the bulk gauge fields \cite{MP,KKS}.
Thus at the compactification scale the theory matches onto a four-dimensional
theory with gaugino masses and negligibly small scalar masses.

Vanishing scalar masses and non-vanishing gaugino masses at a high scale,
as in no-scale models \cite{no-scale}, is very attractive because
evolving the theory to low energies via the renormalization group equation
generates flavor-universal and positive soft scalar masses.
Consistent electroweak symmetry breaking also requires a $\mu$ term of
size comparable to the gaugino masses.
Thus a minimal version of Gaugino Mediation has
only three high energy parameters
\beq
\mu, \quad  M_{1/2}, \quad M_{c} \ .
\label{parameters}
\eeq
Here $M_{1/2}$ is the common gaugino mass at the
unification scale, and $M_c$ is
the compactification scale where the higher dimensional theory is matched
onto the effective four-dimensional theory. Since we wish to preserve the
successful prediction of $\sin^2\theta_w$ from gauge coupling unification
in the MSSM we limit $M_c>M_{GUT}$.

In Section \ref{pheno} of this paper we show that this
scenario, which we call ``Minimal Gaugino Mediation'' (\MgM),
with only the parameters in Eq.~(\ref{parameters}) works very
well phenomenologically. The minimal scenario which we advocate here
differs from the more general models in \cite{KKS,CLNP} in that we do
not introduce soft supersymmetry breaking mass parameters in the
Higgs sector of the theory at $M_c$. Radiative electroweak symmetry breaking
works automatically in \MgM\  and determines $\mu$ by fitting
to the $Z$ mass. Therefore the entire superpartner spectrum of \MgM\ 
can be computed via the renormalization group equations in terms
of only two free parameters: $M_{1/2}$ and $M_c$.
We will see that the running from $M_c$ to $M_{GUT}$ in the
grand unified theory is important for the masses of the
lightest superpartners. We find that the Bino is the LSP and a perfect
cold dark matter candidate in a large region of the models' parameter space. 
``Minimal Gaugino Mediation'' also evades all existing collider
bounds without fine-tuning.

In Section \ref{model} we present a complete and economical model
which gives rise to the \MgM\ boundary condition. The model
generates the hierarchy between the Planck scale and the SUSY breaking scale
with the extra-dimensional dynamical supersymmetry breaking mechanism of
Arkani-Hamed, Hall, Smith, and Weiner \cite{AHSW}. To solve the
$\mu$ problem without introducing a $B\mu$ problem we propose a new
mechanism in which five-dimensional $N=1$ supersymmetry relates $\mu$
and the gaugino mass.

In Section \ref{lastsec} we briefly explain why \MgM\ has no SUSY CP problem,
estimate the neutralino relic density, and conclude. 

%%%%%%%%%%%%%%%%%%%%%%
\section{Sparticle spectrum in \MgM}
\label{pheno}
%%%%%%%%%%%%%%%%%%%%%%
In this section we determine the predictions of \MgM\  for the
spectrum of MSSM particles. The input parameters of the model
are listed in Eq.~(\ref{parameters}). We use the renormalization group
equations (RGEs) of the $\overline{\rm DR}$ scheme to calculate
the soft breaking parameters at the 
electroweak scale. We first outline our procedure for the running
and discuss general features of the evolution. Then we present
the spectrum of superparticles and describe how the experimental limits 
translate into constraints on the parameter space of the model.

At the compactification scale $M_{GUT}\lessim M_c \lessim M_{Planck}/10$
the mass parameters of \MgM\ are
\beq
M_{1/2}\sim \mu \neq 0, \quad m^2=A=B=0 \ .
\label{atMc}
\eeq
We limit the range of compactification scales from below
by the GUT scale in order to preserve the successful prediction
of $\sin^2\theta_w$ from four dimensional unification in the MSSM. Note
that this requirement would still allow compactification scales
slightly below $M_{GUT}$; however as we will discover below, $M_c$ needs to
be slightly larger than $M_{GUT}$ to avoid a charged LSP.
The upper limit on $M_c$ is more model dependent. It
arises from demanding that flavor violating soft masses are sufficiently
small. Such masses are generated from exchange of massive bulk fields with
flavor-violating couplings, which are expected to be present in any fundamental
theory which explains the Yukawa couplings of the Standard Model.
Assuming that the lightest such states have masses of order
$M_{Planck}$ the suppression factor is of the order of $\exp(-M_{Planck}/M_c)$.
Requiring that this exponential suppresses off-diagonal squark masses
sufficiently gives $M_c \lessim M_{Planck}/10$.

To connect the boundary condition of Eq.~(\ref{atMc}) to experiments at the
weak scale we first run from $M_c$ to $M_{GUT}$ in the unified theory and
then run from $M_{GUT}$ to the weak scale with the RGEs of the MSSM.
Gaugino domination, or the no-scale, boundary conditions have been
studied extensively in the literature, however only including renormalization
below the GUT scale~\cite{Castano,BargerBergerOhmann,BGKP}.
Since the renormalization effects above the GUT scale are not discussed
very frequently in the literature we describe them in some detail first.

Naively, one might be tempted to argue against calculating
renormalization effects above the GUT scale because: {\it i.} the running 
above the GUT scale gives only very small masses because
$\log (\frac{M_c}{M_{GUT}}) \ll \log (\frac{M_{GUT}}{M_{weak}})$ and
{\it ii.} the running of soft masses above the GUT scale is model
dependent because the theory above the GUT scale contains new unknown fields
and couplings which enter the RGEs and give rise to unknown threshold effects.
Both of these arguments are invalid as is easy to see:
Argument {\it i.} neglects group theory factors. For example, the mass
which is generated for the right-handed sleptons from running below the
GUT scale is very small because they only couple to hypercharge.
Above the GUT scale, sleptons are unified into larger GUT representations
and the associated larger multiplicity factors more then compensate for
the smaller log.
The second argument would apply in general theories with soft masses, but it
does not apply to \MgM\  where (at one loop) all generated soft masses are
determined by gauge charges only. To understand this consider a
generic one-loop RG equation for scalar soft terms
\beq
\label{schematic}
  \frac{d}{d t} ({\rm soft}) \propto g^2 M_{1/2} + ({\rm soft})
  f(g^2,{\rm SUSY\ couplings})\ .
\label{genericrge}
\eeq
Here the first term is determined entirely by the known gauge charges,
whereas the second term depends on unknown new fields and
couplings. However, in \MgM\ all soft terms for the scalars are zero
at $M_c$. Therefore, the soft masses appearing in the second term are small
(loop-suppressed compared to $M_{1/2}$), and it is a good approximation
to drop the second term. The only remaining model dependence is in
the gauge interactions above the GUT scale.
The predictions depend on the choice of unified gauge group, and we
present predictions for both $SU(5)$ and $SO(10)$.
Furthermore, there is also a weak dependence on the running of the
unified gauge coupling above the GUT scale. We perform our renormalization
group analysis assuming a minimal set of GUT representations
($3\times(10+\bar5)+5+\bar5+24$ for the case of $SU(5)$ and
$3\times16+10+45+16+\bar16$ for $SO(10)$). However, even adding as much as
three additional adjoints to either theory would change the
final scalar masses by at most a few percent.\footnote{A more detailed
discussion of RGEs above the GUT scale can be found in \cite{forthcoming}.}
Finally, note that GUT threshold corrections to the supersymmetry
breaking scalar masses vanish in $\overline{\rm DR}$ so that using the
RGEs gives the complete answer.

Even though we perform our renormalization group analysis numerically
one can also obtain extremely simple approximate formulae for
the soft parameters at the GUT scale as follows. 
At one loop the ratio $\frac{M_{1/2}}{g^2}$ is RGE invariant. Thus,
the running of $M_{1/2}$ is trivial as it traces
the running of the gauge coupling, and we present our results using
$\alpha$ and $M_{1/2}$ evaluated at $M_{GUT}$ rather than at $M_c$.
Assuming that
the running of the couplings above the GUT scale is not too fast
all other soft terms at the GUT scale are then given by
\cite{BarbieriHallStrumia}
\begin{eqnarray}
  A_{top}&= -\frac{2 \alpha}{\pi} M_{1/2}
	  &\ t_c \left[\frac{24}{5},\frac{63}{8}\right]\ ,  \\
  A_{bot}&= -\frac{2 \alpha}{\pi} M_{1/2}
	  &\ t_c \left[\frac{21}{5},\frac{63}{8}\right]\ ,  \\
  B      &= -\frac{2 \alpha}{\pi} M_{1/2}
	  &\ t_c \left[\frac{12}{5},\frac{9}{2}\right]\ , \\
  m^2_{\bf \bar{5}}  &= \frac{2 \alpha}{\pi} M_{1/2}^2
	    & t_c \left[\frac{12}{5},\frac{9}{2}\right]\  , \\
  m^2_{\bf 10}     &= \frac{2 \alpha}{\pi} M_{1/2}^2
		   & t_c  \left[\frac{18}{5},\frac{45}{8}\right]\ ,  
\end{eqnarray}
where $t_c=\log (\frac{M_c}{M_{GUT}})$ ranges between 0 and 4.
Note that we defined the trilinear soft scalar couplings as
$A_{top,bot}\cdot Y_{top,bot}$. All parameters in the equations above are
evaluated at the GUT scale.
The gauge coupling at the unification scale is determined from the low-energy
values of the couplings and it corresponds to $\alpha_{GUT}=1/24.3$.
The first set of numbers in parenthesis applies
to $SU(5)$, the second one to $SO(10)$. With an abuse of notation for the
case of $SO(10)$ we defined $m^2_{\bf \bar{5}}$ to denote the soft mass
for the Higgses of the MSSM, while  $m^2_{\bf 10}$ denotes the common soft
mass of the matter fields.

Below the GUT scale we integrate the one-loop RGEs~\cite{MartinVaughn}
numerically. One loop-running has adequate precision if one uses
the one-loop improved Higgs potential~\cite{HempflingHaber,PBMZ}.
The dominant correction to the lightest Higgs mass comes from
top quark loops below the stop mass threshold. It
can be accounted for by adding the term
\beq
  \frac{3 Y_{top}^4}{16 \pi^2} 
  \log \frac{m_{\tilde{t}_L} m_{\tilde{t}_R}}{m_t^2}
  \left( H_u^\dagger H_u \right)^2 \ .
\label{lutyterm}
\eeq
In addition, we incorporate the contributions
to squark and slepton masses arising from D-terms
as described in Ref.~\cite{BargerBergerOhmann}.

After evolving all soft masses to the weak scale
we impose the constraints which follow from radiative electroweak
symmetry breaking.
This determines the weak scale values of both $\mu$ and $\tan \beta$,
and we are left with only two free parameters:
$M_{1/2}(M_{GUT})$ and the compactification scale $M_c$.
The $\mu$ parameter is multiplicatively renormalized, and it does not
enter any RGE at one loop. Therefore, we will quote its value at the weak
scale.
\begin{figure}[!ht]
\PSbox{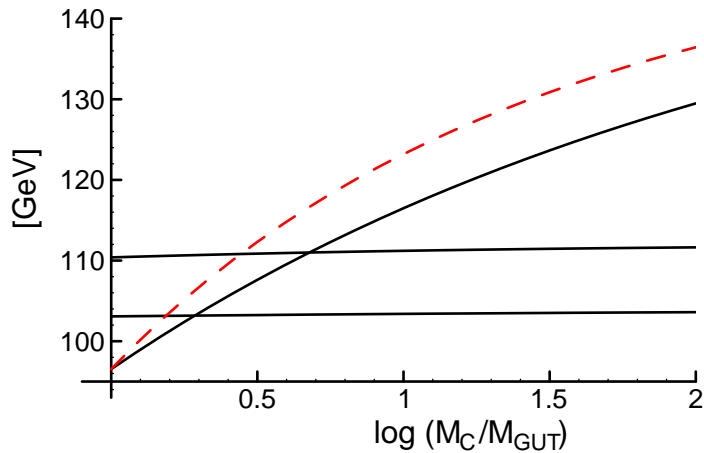 hscale=85 vscale=85 hoffset=45  voffset=20}{13.7cm}{6.5cm}
\caption{The dependence of the masses of the stau, the lightest neutralino
and the lightest Higgs on the compactification scale for fixed
$M_{1/2}=250$~GeV. The neutralino mass ($104$ GeV) and the Higgs
mass ($111$ GeV) are almost independent of the compactification scale
and the GUT gauge group. The rising solid line and the dashed line
indicate the mass of the stau in $SU(5)$ and $SO(10)$, respectively.}
\label{fig:aboveGUT}
\end{figure}

Figure \ref{fig:aboveGUT} illustrates the significance of
the RG evolution above the GUT scale. Without running above the GUT scale
($M_c=M_{GUT}$) the stau is the LSP; however for any compactification scale
larger than only $1.5 M_{GUT}$ the stau is heavier than the lightest
neutralino.
Note that the dependence of the stau mass on $t_c=\log(\frac{M_c}{M_{GUT}})$
is stronger in $SO(10)$ than in $SU(5)$. This follows from the
larger group theoretical factors in $SO(10)$ which cause soft masses
above the GUT scale to be generated more efficiently.

\begin{figure}[t]
 \PSbox{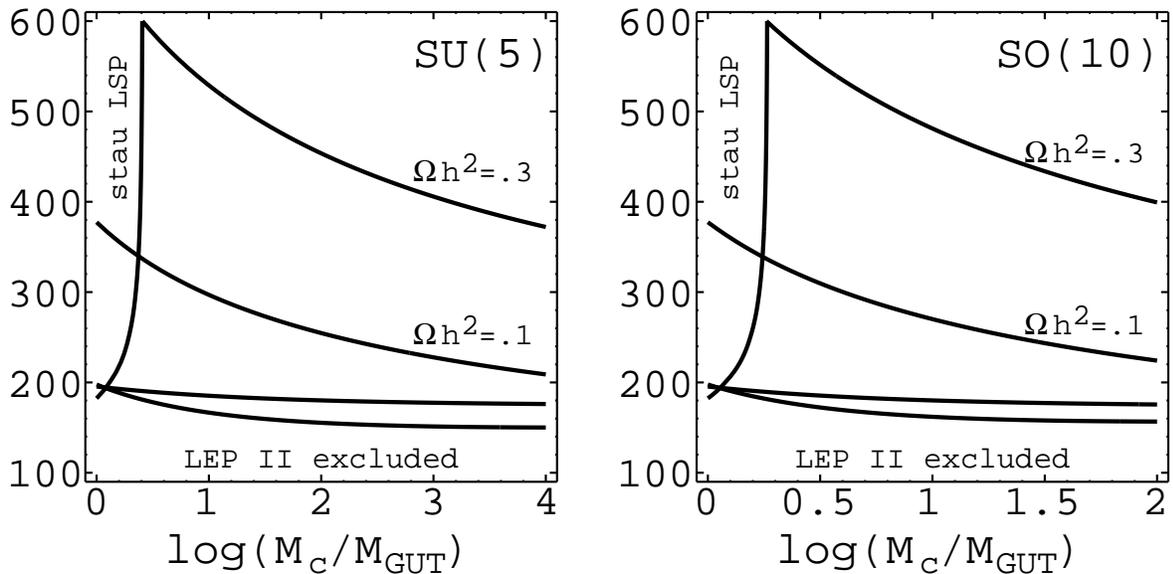 hscale=80 vscale=80 hoffset=-10  voffset=0}{13.7cm}{7cm}
 \caption{The allowed region of the parameter space for \MgM\ with
 $SU(5)$ and $SO(10)$ unified group. The curves at the bottom of the
 figure correspond to LEP II limits on the masses of the Higgs
(more restrictive)  and stau (less restrictive).
Demanding that relic LSPs contribute the cosmologically
 preferred amount of cold dark matter and do not over-close the universe
 singles out the region of parameter space between the lines labeled
 $\Omega h^2 = 0.1$ and $0.3$.}
\label{fig:exclusion}
\end{figure}

The allowed parameter space for $SU(5)$ and $SO(10)$ \MgM\ models is
presented in Figure~\ref{fig:exclusion}. We find a lower bound on $t_c$
from requiring that the LSP be neutral. An upper bound on $t_c$ is not
shown on the figure, but as discussed above, flavor violating
effects due to massive bulk fields limit $t_c\lsim 4$.
Since $M_{1/2}$ is the only source for superpartner masses, the
experimental lower limits on superpartner masses and the Higgs mass
translate into lower limits on $M_{1/2}$. 
In particular, we find that the LEP II limits~\cite{LEP2} on the Higgs mass
($m_{h^o}\geq$ 106 GeV)\footnote{The Standard Model Higgs bound rather than
the much weaker SUSY Higgs bound applies in the entire
allowed parameter space because $\mu$ is sufficiently large so that
the heavier Higgs fields decouple and the production cross section
becomes Standard-Model-like.} and the right handed slepton masses
($m_{\tilde{\tau}}\geq 75$ GeV and $m_{\tilde{e}}\geq 95$ GeV)
imply $M_{1/2}\gsim 180$ GeV. Furthermore, we find that the $\mu$ parameter
in our model is given by $\mu = 3/2 M_{1/2}$ to an accuracy of better
than 2$\%$ for all values of $t_c$. This implies a lower bound
$\mu \gsim 270$ GeV with an associated mild tuning of the $Z$ mass.
The figure also shows contours of the relic abundance of the
lightest neutralino corresponding to $\Omega_{\chi}h^2=0.1,\ 0.3$.
The LSP relic abundance calculation is particularly simple in our model,
we discuss it briefly in Section \ref{lastsec}.

\begin{figure}[!bht]
\PSbox{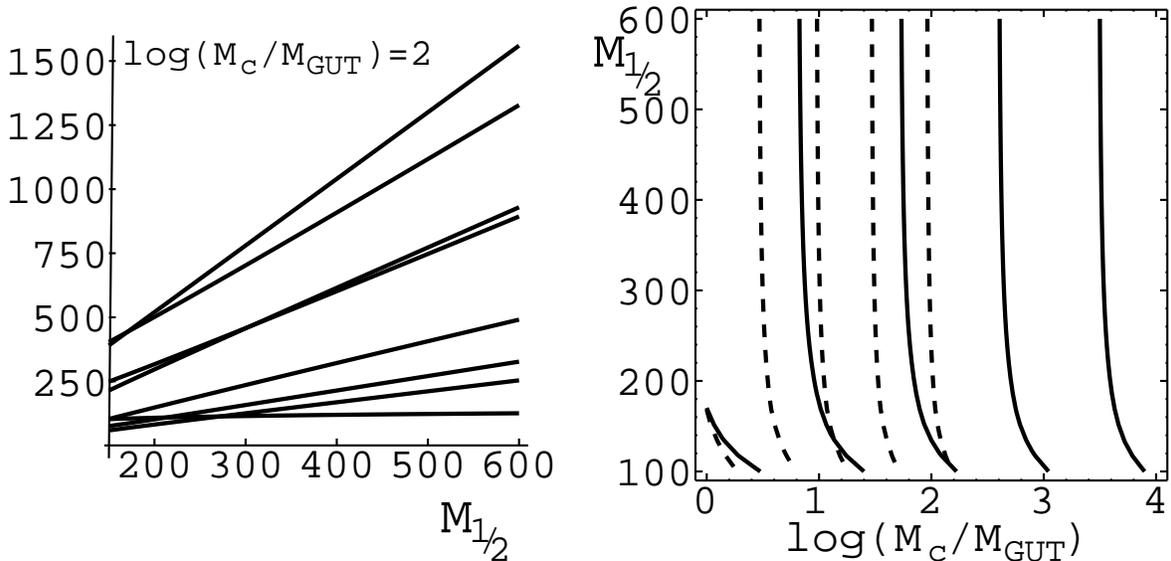 hscale=80 vscale=80 hoffset=0  voffset=0}{13.7cm}{7cm}
 \caption{Left graph: particle masses as functions of $M_{1/2}$ for fixed
 compactification scale. The lines correspond to (from lightest to heaviest 
 at $M_{1/2}=600$~GeV) the lightest Higgs, lightest neutralino,
 right-handed stau, second lightest neutralino, heavy chargino,
 pseudoscalar Higgs, left-handed stop, and gluino.
 Right graph: contours of constant $\tan \beta$. The contours correspond to
 $\tan \beta$ of 9, 12, 15, 18, and 21 from left to right. The solid lines
 are for $SU(5)$, the dashed ones for $SO(10)$.}
 \label{fig:masstan}
\end{figure}
Figure~\ref{fig:masstan} shows the \MgM\ spectrum as a function of
the gaugino mass for the example case of an $SU(5)$ GUT with
$\log (\frac{M_c}{M_{GUT}})=2$.
The qualitative features of the spectrum are generic and do not depend
on the choice of grand unified group or compactification scale.
The masses of all superpartners and Higgs fields, except for the
lightest Higgs, rise linearly with $M_{1/2}$. 
As in minimal supergravity the LSP is a Bino-like neutralino.
The right-handed stau is the next-to-LSP. As usual, colored superpartners
are heaviest, followed by charginos, neutralinos and Higgses with masses
of order $\mu$. 
The mass of the lightest Higgs particle increases only
logarithmically with $M_{1/2}$ through the one-loop
improvement of the Higgs potential as described in Eq.~(\ref{lutyterm}).

Figure~\ref{fig:masstan} also shows contours of constant $\tan \beta$
in the $M_{1/2}-t_c$ plane. Note that in the allowed region $\tan \beta$
is almost independent of $M_{1/2}$, but it increases with $t_c$.

Since \MgM\ has only 2 free parameters, measuring the masses of only
two particles is in principle sufficient to determine the input
parameters and predict the entire superpartner spectrum.
In practice, presumably the Higgs will be the first new
particle to be discovered. This is because the MSSM Higgs mass bound of
130~GeV applies also to the \MgM\ Higgs which could therefore
be discovered (or ruled out) at Run II of the Tevatron~\cite{fnalHiggs},
and might even be seen at LEP 205. 
The mass of the Higgs would give an estimate of $M_{1/2}$. Should
$M_{1/2}$ be close to 200~GeV, there is a chance that LEP or
the Tevatron will discover the first superpartners. For 
low enough compactification scales LEP would find the right-handed stau
and/or selectron. Independent of the compactification scale the
Tevatron could then observe charginos in the tri-lepton channel
\cite{sugrafnal}. For larger $M_{1/2}$ we would have to wait for
the LHC.

It is exciting that observation of the first superpartner immediately
also leads to a first test of the model. This is because the discovery
would allow a mass measurement of both the discovered superpartner
as well as the LSP mass from the distribution of the missing energy.
One could then use the measured masses of the Higgs and Bino to obtain
two independent determinations of $M_{1/2}$ and therefore test the model.
Once we know a mass of any of the sleptons we can extract
the remaining free parameter -- $t_c$. Note that discovery of just a few
of the lightest superpartners would already allow a determination of
the GUT gauge group!

%%%%%%%%%%%%%%%%%%%%%%%%
\section{\MgM, an explicit model}
\label{model}
%%%%%%%%%%%%%%%%%%%%%%%%%

In this section we describe a simple model which breaks supersymmetry
and yields only gaugino masses at the compactification scale. Our model
is complete: it generates exponentially small supersymmetry breaking which
is mediated to the gauginos via a higher dimensional operator, and $\mu$
is naturally of the same order as the gaugino masses. The model
combines the idea of ``gaugino mediation'' with the supersymmetry breaking
mechanism proposed in \cite{AHSW}.

\begin{figure}[!ht]
\PSbox{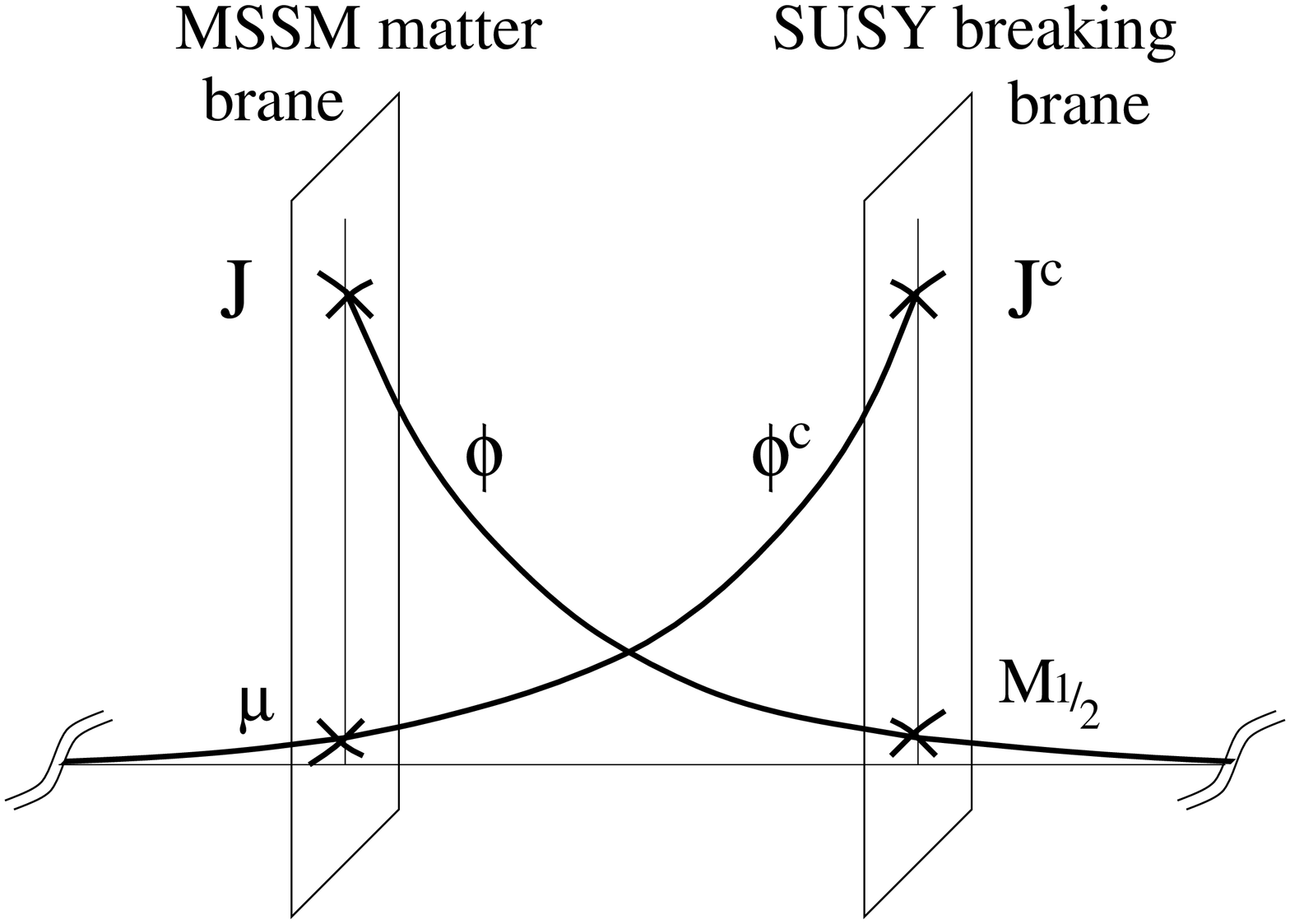 hscale=50 vscale=40 hoffset=50  voffset=10}{13.7cm}{7cm}
\caption{A brane configuration which leads to the Minimal Gaugino Mediation
boundary condition. Pictured is the extra
dimension from left to right with periodic boundary conditions.
We also show the exponentially decaying vacuum expectation values
of $\phi$ and $\phi^c$ which are responsible for generating hierarchically
small supersymmetry breaking and the $\mu$ term. }
\label{golden-gate-bridge}
\end{figure}

To begin we recall the higher dimensional set-up of \cite{KKS}. The
MSSM matter and Higgs fields live on a 3+1 dimensional brane embedded in
one extra dimension. Supersymmetry is broken dynamically on a parallel 
brane which is a distance $L$ apart from the matter brane
(Fig. \ref{golden-gate-bridge}). The MSSM gauge fields and
gauginos live in the bulk of the extra dimension. We take
this extra dimension to be circular, with radius $R$. In order to
preserve the quantitative prediction of $\sin^2 \theta_w$ from gauge
coupling unification in the four dimensional MSSM, we demand that the
compactification scale be higher than the GUT scale,
$R^{-1}\equiv M_c \ge M_{GUT}$.\footnote{A similar constraint on the
compactification scale also follows from demanding that the
extra-dimensional theory remains perturbative up to the five dimensional
Planck scale $M$,\ 
$g_{GUT}^2={g_5^2\over 2\pi R} < {24 \pi^{5/2} \over 2 \pi R M}$. Using
$2 \pi R M^3 = M_{Planck}^2$ this becomes $R M_{Planck} < 750$.}

%%%%%%%%%%%%%%%%%%%%%%%%%%%%%%%%%%%%%%%%%%%%%%%%%%%%%%%%%%%%%%%%%%%%%
%%%%%%%%%%%%%%%%%%%%%%%%%%%%%%%%%%%%%%%%%%%%%%%%%%%%%%%%%%%%%%%%%%%%%

In our model, supersymmetry breaking manifests
itself in a vacuum expectation value for the F-component, $X_F$, of a chiral
superfield $X$ on the SUSY breaking brane. The MSSM gaugino
fields can couple to $X$ directly, giving a gaugino mass
\beq
\label{gaugmass}
\int dx_5 \delta(x_5-L) \int d^2\theta \ {X W W \over M^2} \rightarrow 
     { X_F \over V M^2} \lambda \lambda \ .
\eeq
Here, the factor of the extra-dimensional volume ($V=2\pi R$ for a circle)
arises from the wave function normalizations of the four-dimensional
gaugino fields $\lambda$.
All other soft supersymmetry breaking parameters in the MSSM, such as soft
scalar masses $X^\dagger XQ^\dagger Q$, are suppressed at short distances
by extra-dimensional locality \cite{KKS}. The low-energy
values of these parameters are
generated from the renormalization group equations as discussed in the
previous section.

It is useful to discuss the exact form of the
short-distance suppressions in more detail. In general, there are two
possible sources for such terms: direct contact terms suppressed by
the cut-off $M$ or non-local terms from loops of the light bulk gauge fields.
The contact terms are not present in the effective theory below $M$ to all
orders in the local expansion in inverse powers of $M$ because they connect
fields at different positions. However, this does not
preclude the appearance of terms with coefficients $e^{-LM}$ which do
not have an expansion in local operators. These operators are expected
to be flavor violating and are therefore strongly
constrained experimentally \cite{CPconstraints}.
The most stringent constraint comes from CP violation in the K system
and gives roughly $e^{-LM}\lessim 10^{-4}$ or $LM \gtrsim 8$.
Therefore, the allowed range for the compactification scale is
$M_{GUT} \lessim M_c \lessim M_{Planck}/10$.

The other source of short-distance scalar masses -- loops of bulk gauge
fields -- leads to finite contributions to the masses which are suppressed
by additional powers of the separation $L$ relative to the gaugino mass
(\ref{gaugmass}). However, they are flavor universal because
they arise from gauge interactions. As discussed in detail in \cite{KKS}
these contributions are negligible compared to the much larger contributions
from the renormalization group evolution.
  
In the following Subsections we turn to discussing the mechanism of
supersymmetry breaking and the origin of the $\mu$ term in the model.
Our mechanism for breaking supersymmetry and stabilizing the radius
of the extra dimension is taken directly from the elegant paper of
Arkani-Hamed {\it et. al.} \cite{AHSW}. In the following, we summarize
their discussion and apply it to our model. Our solution to the $\mu$
problem is new.

\subsection{Supersymmetry breaking}

Following \cite{AHSW}, we keep track of four-dimensional
$N=1$ supersymmetry by employing four-dimensional $N=1$ superspace notation
and treating the $x_5$ coordinate as a label.
The action for a massive five-dimensional hypermultiplet $(\Phi,\Phi^c)$
then reads
\beq
\int d^4x dx_5 
    \left( \int d^4\theta (\Phi^\dagger \Phi + \Phi^{c\dagger} \Phi^c)
          +\int d^2\theta \Phi^c (m+\partial_5) \Phi \right)\ .
\label{bulkaction}
\eeq
The advantage of this formalism is that it is straightforward to write down
$N=1$ supersymmetric couplings of $\Phi$ to boundary fields. 
The supersymmetry breaking model of \cite{AHSW} consists of the bulk
field $\Phi$ with superpotential couplings to a source $J$
and a field $X$ which are localized on different branes\@.
$J$ is localized on the matter brane at $x_5=0$, while $X$ on
the SUSY-breaking brane at $x_5=L$
\beq
\int dx_5 \left(- \delta(x_5) \sqrt{M} J \Phi^c +
                  \delta(x_5-L) \sqrt{M} X \Phi \right) \ .
\label{susybreaking}
\eeq
Here we have suppressed coupling constants but inserted
factors of the fundamental mass scale $M$ to keep track of mass
dimensions. The vacuum equations for the scalar field are then
\bea
\label{eom1}
\Phi_F &=& \delta(x_5-L) \sqrt{M} X + (m-\partial_5)\phi^c = 0, \\
\Phi^c_F &=& - \delta(x_5) \sqrt{M} J + (m+\partial_5) \phi = 0 \ .
\eea
On a circle $x_5 \in [0,2\pi R)$ the equation for $\Phi$ has the unique
solution
\beq
\phi = \sqrt{M} J\ {e^{-mx_5} \over 1 - e^{-m 2 \pi R}} \ .
\label{shineeq}
\eeq
Thus the source $J$ ``shines'' a vacuum expectation value for the bulk
scalar $\phi$ which decays exponentially with increasing $x_5$
(see Fig. \ref{golden-gate-bridge}).
Supersymmetry is broken because $X$ obtains a non-vanishing $F$-component
\beq
\label{Xeom}
X_F = \sqrt{M} \phi(L) = M J\ {e^{-mL} \over 1 - e^{-m 2 \pi R}}
             \sim M J\ e^{-mL}\ .
\eeq
Assuming a source $J \sim M$ and
a mass $m \sim M$ one finds $X_F \sim M^2 \ e^{-ML}$.

Note that this model is a higher dimensional generalization of a simple 
O'Raifeartaigh model. The source $J$ forces a non-zero expectation
value for the field $\phi$, that is in conflict with the $X$ equation
of motion which requires $\phi=0$. The role of the extra dimension is
to modulate the resulting supersymmetry breaking by the factor $e^{-ML}$.
Coupling the field $X$ to the gauge fields as in Eq. (\ref{gaugmass}) then
results in non-vanishing gaugino masses
\beq
M_{1/2}={X_F \over 2\pi RM^2}\sim {J\over 2\pi RM} e^{-ML}\ .
\label{gaugmas}
\eeq
As in ordinary O'Raifeartaigh models, the scalar expectation value of $X$
is undetermined classically. A non-vanishing expectation
value can be seen to act as a source for $\Phi^c$ from Eq. (\ref{eom1}).
In order to simplify the analysis we assume that the $X$-expectation
value is zero. This may either be enforced by additional tree-level
superpotential terms on the supersymmetry breaking brane such as
$ \delta(x_5-L) [ X Y + Y^2 Z ]$, or it could be a result of quantum
corrections lifting the flat direction.

\subsection{The $\mu$ term}

To generate a $\mu$ term of the correct size we utilize the
$\phi^c$ component of the
superfield $(\Phi,\Phi^c)$. To break supersymmetry we used an
expectation value for $\phi$ which was ``shining''
clockwise from the source $J$ on the matter brane towards the
supersymmetry breaking brane. For generating $\mu$ we ``shine'' an
expectation value for $\phi^c$ by adding the superpotential
\beq
\int dx_5 \left(- \delta(x_5-L) \sqrt{M} J^c \Phi +
                 \delta(x_5) {\kappa \over \sqrt{M}} \Phi^c H_u H_d \right) \ .
\label{musuperpot}
\eeq
The new terms modify the equations of motion
\bea
\label{eom2}
\Phi_F &=&- \delta(x_5-L) \sqrt{M} J^c + (m-\partial_5)\phi^c = 0, \\
\Phi^c_F &=&- \delta(x_5) \sqrt{M} J + (m+\partial_5) \phi = 0 \ ,
\eea
where we have assumed
that the vacuum expectation values of $X$ and $H_u H_d$ are negligible
compared to $J, J^c \sim M$. As mentioned in the previous Subsection this
can be enforced by adding suitable brane potentials.

We see that the $\phi$ equation is unchanged, while the new source $J^c$
also ``shines'' an expectation value for $\phi^c$
\beq
\phi^c = \sqrt{M} J^c\ 
      {e^{m(x_5-L)-m2\pi R \, \theta(x_5-L)} \over 1 - e^{-m 2 \pi R}} \ .
\eeq
Note that since we have placed the source on the supersymmetry
breaking brane $\phi^c$ is ``shined'' in the opposite direction
from $\phi$, as depicted in Fig.~\ref{golden-gate-bridge}.
The generated $\mu$ term is equal to
\beq
\mu={\kappa \over \sqrt{M}} \phi^c(0)
   =\kappa J^c\ {e^{-mL} \over 1- e^{-m2\pi R}}
      \sim \kappa J^c\ e^{-mL}\ .
\eeq
Comparing this to the gaugino mass Eq. (\ref{gaugmas}) we find that we
need to set $\kappa \sim 1/(2\pi RM)\sim 1/100$.

Note that $\mu$ has the exact same exponential suppression factor
$e^{-mL}$ as $m_{1/2}$. This follows from the fact that $\phi$ and $\phi^c$
reside in the same five dimensional supersymmetry multiplet. In other words, five
dimensional supersymmetry relates the exponential suppression factors
appearing in $\mu$ and $M_{1/2}$.
It is disappointing that because of the volume suppression in the
gaugino masses we still need to choose a small coupling $\kappa$ to
get $\mu \sim M_{1/2}$.
However, $\kappa$ is a superpotential coupling and as such
can be small naturally. Note that the spatial separation
of the supersymmetry breaking $X_F$
from the location of the Higgs fields does not allow a $B\mu$
term at the high scale. We therefore do not have the usual problem
$B \sim 16 \pi \mu$ which haunts most other approaches to the $\mu$ problem.
Finally, we emphasize that this new extra-dimensional solution to the $\mu$
problem does have broader applicability.

\subsection{Radius stabilization}

In the discussion above we have assumed that the radius $R$ of the extra
dimension and the distance $L$ between the branes are fixed. In a complete
theory both parameters correspond to fields. We now discuss a simple
supersymmetry-preserving mechanism to stabilize both $R$ and $L$.
Our mechanism is a trivial modification of \cite{AHSW}.
In its simplest form it requires a single additional massive bulk
hypermultiplet $(\Psi,\Psi^c)$ with couplings to brane fields
\beq
\int dx_5 \left( -\delta(x_5) \sqrt{M} [I \Psi^c + I^c \Psi] +
        \delta(x_5-L) \sqrt{M} [A (\Psi-\Lambda \sqrt{M}) +                    
                               A^c (\Psi^c-\Lambda^c \sqrt{M})] \right)\ .
\eeq
Assuming that the brane fields $A,A^c$ have no vacuum expectation
values\footnote{In the absence of supersymmetry breaking these expectation
values are flat directions. It is straightforward to enforce the vanishing
expectation values, for example by adding a brane superpotential
$\delta(x_5-L) [A B + B^2 C]$ for $A$ and similarly for $A^c$.}
one finds the following equations of motion
\bea
\Psi_F &=& - \delta(x_5) \sqrt{M} I^c + (m_\Psi-\partial_5)\psi^c = 0,\\
\Psi^c_F &=& - \delta(x_5) \sqrt{M} I + (m_\Psi+\partial_5) \psi = 0, \\
A_F&=&\psi(L)-\Lambda \sqrt{M}=0\ ,\quad A^c_F=\psi^c(L)-\Lambda^c \sqrt{M}=0,
\eea
which have unique supersymmetry preserving solutions for $R$ and $L$.
For example for symmetric values of the parameters $\Lambda=\Lambda^c$
and $I=I^c$ we find
\beq
L=\pi R ={1\over m_\Psi} {\rm arcsinh}\left(I \over 2 \Lambda \right) \ .
\eeq
Thus, for $I$ and $\Lambda$ of order $M$ a radius of the desired size
is generated by choosing a relatively small mass for the bulk scalar
$m_\Psi \sim M/30$.

%%%%%%%%%%%%%%%%%%%%%%%%
\section{Discussion}
\label{lastsec}
%%%%%%%%%%%%%%%%%%%%%%%%

Minimal Gaugino Mediation is a very compelling and predictive theoretical
framework which solves all supersymmetric naturalness problems without
fine-tuning. 

\MgM\ solves the supersymmetric flavor problem: At the high scale $M_c$
the scalar masses and A-terms vanish, and therefore the only flavor violation
in renormalizable couplings resides in the Yukawa couplings.
Gaugino loops generate universal positive scalar masses at low energies.
Small non-universalities in the masses arise from the Yukawa interactions,
but these contributions do not lead to new flavor violation because they are
aligned with the Yukawa matrices. An exception to this is the running of
the scalar masses above the GUT scale where flavor is broken by unified
interactions~\cite{BarbieriHallStrumia,aboveGUT}.
Since the right-handed sleptons are light in \MgM\ event rates for lepton
flavor violating processes such as $\mu\rightarrow e \gamma$ might be
near the experimental bounds.

\MgM\ solves the supersymmetric CP problem: This
is easy to understand by realizing that at the compactification scale
(where $m^2=A=B=0$) the phases in $M_{1/2}$ and $\mu$ 
can be removed by phase redefinitions of the gaugino fields and the
Higgs superfields, respectively. Therefore, the theory
has no new phases beyond the phases in the Yukawa couplings
and no supersymmetric CP violation. 
This does not solve the strong CP problem however. 

\MgM\ is very predictive and therefore testable: The model has only two
free parameters which implies that there are many relations between the 
masses of the superpartners and Higgses which can be tested experimentally.

\MgM\ has a great cold dark matter candidate: The LSP of \MgM\ is almost
a pure Bino for most of the parameter space. This makes the calculation of
the relic neutralino (Bino) density relatively easy, because in this scenario
neutralino annihilations are dominated by the t-channel exchange of the
right-handed sleptons. If one ignores
the small (but interesting) region of parameter space where the stau
and neutralino are degenerate to within 5$\%$ (and where co-annihilations
are important~\cite{Toby}) the relic neutralino abundance is
given by \cite{Jimandothers}
\beq
\Omega_\chi h^2 \approx {(m^2_{\tilde{l}_R}+m^2_\chi)^4 \over 
                   (1.4\ TeV)^2 m^2_\chi
             (m^4_{\tilde{l}_R}+m^4_\chi)}
            \approx {m^2_{\tilde{l}_R} \over (480\ GeV)^2} \ .
\label{abundance}
\eeq
This formula is accurate to about 20$\%$ over the whole parameter space
plotted in Figure \ref{fig:exclusion} except for where neutralinos and 
staus are almost degenerate (a narrow band surrounding the ``stau LSP''
excluded regions). For \MgM\ Eq.~(\ref{abundance}) yields abundances which
generically are cosmologically safe and often lie
within the cosmologically interesting regime $0.1 < \Omega_\chi h^2 < 0.3$
as is evident from Figure \ref{fig:exclusion}.

\MgM\ is theoretically well motivated: Separation of SUSY breaking and
the MSSM matter fields onto two different branes naturally gives rise
to the Gaugino Mediation boundary condition. If the Higgs fields also live
on the MSSM matter brane then all supersymmetry breaking soft Higgs mass
parameters vanish, giving \MgM. The model is very economical and unifies.
We believe that the model is sufficiently ``conservative'', successful in
solving all the problems of supersymmetry, and elegant that it has a real
chance of describing Nature.

%%%%%%%%%%%%%%%%%%%%%%%%
\acknowledgements
%%%%%%%%%%%%%%%%%%%%%%%%

We thank Nima Arkani-Hamed, Howard Baer, Howie Haber, Markus Luty,
Chris Kolda, Graham Kribs, Kirill Melnikov, Michael Peskin, and Jim Wells for
useful discussions. MS is supported by the DOE under contract
DE-AC03-76SF00515. WS is supported by the DOE
under contract DOE-FG03-97ER40506.

\def\pl#1#2#3{{\it Phys. Lett. }{\bf B#1~}(19#2)~#3}
\def\zp#1#2#3{{\it Z. Phys. }{\bf C#1~}(19#2)~#3}
\def\prl#1#2#3{{\it Phys. Rev. Lett. }{\bf #1~}(19#2)~#3}
\def\rmp#1#2#3{{\it Rev. Mod. Phys. }{\bf #1~}(19#2)~#3}
\def\prep#1#2#3{{\it Phys. Rep. }{\bf #1~}(19#2)~#3}
\def\pr#1#2#3{{\it Phys. Rev. }{\bf D#1~}(19#2)~#3}
\def\np#1#2#3{{\it Nucl. Phys. }{\bf B#1~}(19#2)~#3}
\def\xxx#1{{\tt [#1]}}

\end{document}